\documentclass[11pt,a4paper]{article}
\pdfoutput=1
\usepackage{jcappub}
\usepackage{epsfig}
\usepackage{amsmath}
\usepackage{subfigure}

\DeclareSymbolFont{AMSa}{U}{msa}{m}{n}
\DeclareSymbolFont{AMSb}{U}{msb}{m}{n}
\let\Box\relax
\DeclareMathSymbol{\Box}{\mathord}{AMSa}{"03}

\def\cR{\mathcal{R}}
\def\cH{\mathcal{H}}
\def\cP{\mathcal{P}}
\def\abi{a_i}

\def\e{\eta}
\def\wbi{w_i}
\def\wi{w_0}

\def\Re{\mbox{Re}\,}
\def\Im{\mbox{Im}\,}

\newcommand{\be}{\begin{equation}}
\newcommand{\ee}{\end{equation}}
\newcommand{\bea}{\begin{eqnarray}}
\newcommand{\eea}{\end{eqnarray}}

\subheader{\begin{flushright}
UTTG-14-11\\
TCC-021-11
\end{flushright}}

\title{The Inflationary Wavefunction and its Initial Conditions}

\author[a,b]{Daniel Carney,}
\author[a,b]{ Willy Fischler,}
\author[a,b]{ Sonia Paban}
\author[a,b,c]{ and Navin Sivanandam}

\affiliation[a]{Texas Cosmology Center, University of Texas, Austin, TX 78712}
\affiliation[b]{Theory Group, Department of Physics, University of Texas, Austin, TX 78712}
\affiliation[c]{African Institute for Mathematical Sciences, Muizenberg, Cape Town, SA}

\emailAdd{carney@physics.utexas.edu}
\emailAdd{fischler@physics.utexas.edu}
\emailAdd{paban@physics.utexas.edu}
\emailAdd{navin.sivanandam@gmail.com}

\abstract{We explore the effect of initial conditions on the inflationary wavefunction and their consequences for the observed spectrum of primordial fluctuations. In a class of models with a sudden transition into inflation we find that, for a reasonable set of assumptions about the reheat temperature and the number of e-foldings, it is possible for initial conditions set by a pre-inflationary epoch to have an observable effect.}


\arxivnumber{1109.6566}

\begin{document}

\maketitle

\section{Introduction}
An apparent success of inflation is that the spectrum of fluctuations is insensitive to initial conditions. Such an expectation may be well grounded for the case of an inflationary era with a large number of e-foldings followed by a finite deceleration era.  We will show in this paper, however,  that for a reasonable set of assumptions on the number of e-foldings and reheat temperature, the wavefunction of the inflaton at the onset of inflation can impact the formula for the spectrum of fluctuations.

The initial wavefunction for the inflaton is not without constraints, as explored in a number of papers, including \cite{Eboli:1988qi, Kaloper:2002cs, Anderson:2000wx,Destri:2009hn}. The wavefunction dependence on modes with large momenta, i.e. momenta of the order of a UV cut-off is pretty much fixed.  The expectation value of the energy-momentum tensor is cut-off dependent, restricting the state for high momenta modes to be consistent with renormalizability. This requires that $\langle T_{\mu \nu}\rangle$  is a sum of terms obtained from the variation of generally covariant and renormalizable terms in the action. Moreover, since shorter length scales are stretched as inflation progresses, the initially high momenta modes can be observationally relevant. Thus, for sufficient numbers of e-foldings, the initial conditions for modes of interest are fixed.

In this paper, we will work with a fixed UV cut-off for physical momenta below the fundamental scale thereby avoiding speculations about trans-Planckian physics\footnote{We should note that the adiabatic approximation is debatable as the cut-off nears the Planck scale. Indeed, as the UV-IR connection sets in, the adiabatic approximation may fail. We should therefore choose the cut-off a safe distance away from the Planck scale.}. This implies that as FRW time evolves, ``new'' coordinate momenta corresponding to physical momenta below the fixed UV cut-off ``emerge''. The wavefunction dependence on modes as they emerge below the cut-off with passing FRW time will be subject to renormalizability constraints.
Although existing work along the lines discussed above would seem to suggest that there is little hope of observing the effects of the initial inflationary state, we find that this is not strictly true. In particular the dependence of the wavefunction on modes with momenta much smaller than the UV cut-off but larger than the Hubble parameter at the onset of inflation will be determined by their history prior to the inflationary epoch. If such modes are observationally relevant then we must learn about the constraints (if any) on them.
To study the possible consequences of the scenario of the previous paragraph we focus on a particular model of the pre-inflationary universe. We use a kinetic energy dominated scalar field cosmology to give us a decelerating expansion before the start of inflation. In such a cosmology there are again UV constraints that set the initial conditions for modes of sufficient high wavenumber (as above these arise from requiring that the renormalized expectation value of energy momentum tensor is generally covariant). However, these constraints are different from those that apply for an accelerated expansion, and the differences can affect observable quantities.

We find that if the transition from the kinetic energy dominated era to the inflationary era is sudden, the expressions for the spectral properties of the primordial fluctuations are modified, particularly the dependence of the ratio $\delta\rho\over{\rho}$ on the slow roll parameter $\epsilon$, is different. Rather than the usual $1/\epsilon$ scaling, the modified scalar spectrum has $1/\epsilon^2$. Furthermore, we find a modulation in momentum dependance of the new spectrum, albeit one too fast to be observed except in average. These differences and their consequences are the main results of this paper.

Before moving on we note that effect of sudden changes in the inflationary potential has been studied previously in \cite{Starobinsky:1992ts,Leach:2001zf,Adams:2001vc}. As in our work, the sudden change affects the normalization of the power spectrum and can lead to an enhancement or oscillations for some range of $k$. While the underlying mathematical mechanism is similar, the authors in these papers are concerned with temporary violations of slow roll during inflation, rather than the effect of a sudden transition from a pre-inflationary epoch on the initial wavefunction.

The paper is organized as follows. In section \ref{schroedpicture} we briefly summarize our method of solving the Scr\"{o}edinger equation and the class of wavefunctions we will be concerned with. Then, in section \ref{initialcond} we discuss how UV physics constrains the form of the wavefunction. Section \ref{modelevol} describes the particular model we have studied and the evolution of that model. We explain the possible observable consequences of the model in section \ref{observables}, highlighting how deviations from the standard vacuum may be observed. Finally, we offer our conclusions in section \ref{conclusions}.

\section{Finding The Power Spectrum in the Schr\"{o}edinger Picture}\label{schroedpicture}
We work with FRW time and a mostly plus signature, so that the spacetime we study is described by the usual Robertson-Walker metric:
\be
ds^2=-dt^2+a^2(t)d\mathbf{x}^2\, .
\ee
We wish to study the power spectrum in a  model  of single field inflation. To this end we will parameterize the scalar fluctuations in terms of the comoving curvature perturbation, $\cR$. This quantity  is gauge invariant and conserved  outside the horizon \cite{Mukhanov:1990me}. The power spectrum for $\cR$ is
\be
\langle\cR_{\bf k} \, \cR_{\bf k'}\rangle = \left(2 \pi\right)^3\delta\left({\bf{k + k '}}\right)P_{\cR}\left(k\right)\, ,
\ee
where $\cR_{\bf k}=\int d^3x\,\cR(x) e^{-i\mathbf{k}\cdot\mathbf{x}}$.

The action for $\cR$, as derived in \cite{Maldacena:2002vr}, is
\begin{equation}\label{ham}
S= \frac{1}{2} \int dt\,d^3x\,a^3(t) \frac{\dot{\phi}^2}{H^2}\left[\dot{\cR}^2-\frac{1}{a^2}\left({\partial \cR}\right)^2\right]\, .
\end{equation}
$\phi$ is the background inflaton field and $H\equiv\dot{a}/{a}$. When written in terms of the variable $\cR_{\bf k}$, this action corresponds, for every value of $k$, to a simple harmonic oscillator with a time-dependent mass and frequency. Its Hamiltonian is
\begin{equation}\label{hoe}
\mathcal H = \frac{1}{2 m(t)} \,\, \Pi_{\cR}^2 + \frac{1}{2} m(t) \,\, \omega^2(t) \cR^2\, ,
\end{equation}
where
\begin{equation}\label{mfd}
m(t) \equiv \frac{ a^3(t) \dot{\phi}^2(t)}{H^2(t)} \hspace{4ex} \mbox{and} \hspace{4ex} \omega(t) = \frac{k}{a(t)}\, .
\end{equation}

\subsection{The Schr\"{o}dinger Picture}
Although it is more conventional to work in the Heisenberg picture when analysing inflationary perturbations, in this work we shall use the Schr\"{o}dinger picture (see for example \cite{Eboli:1988qi, Guven:1998, Fischler:1985ky}). There are two principle reasons for this. The first is aesthetic, we feel that when viewing the problem of inflationary perturbations in the framework of the evolution of some specified initial state, the Schr\"{o}dinger picture provides a more useful physical picture. More important then aesthetic considerations, though, are the technical issues that arise when considering quantum fluctuations for a scalar potential with a sharp discontinuity -- this sort of potential is the basis of the model we shall consider presently.

When studying gravitational systems with sharp transitions between one region of spacetime to another, one needs to impose the appropriate junction conditions. For the case small perturbations during a sudden change in the cosmology, the problem has been studied by Mukhanov and Deruelle \cite{Deruelle:1995kd}, who adapt the known Israel conditions to cosmological perturbations. In the longitudinal (conformal Newtonian) gauge, the appropriate conditions, for the sharp transition used in the paper,  are:
\begin{align}
\left[\Phi_l\right]_{\pm}&=0\, ,\nonumber\\
\left[\cR \right]_{\pm}&=0\, .
\end{align}
$\Phi_l$ is the Newtonian potential in the longitudinal gauge, $\cR$ is the comoving curvature perturbation. Requiring that the perturbations in the Heisenberg picture satisfy this matching conditions is equivalent to the Schr\"{o}dinger formalism used throughout  the paper.

Before moving on to finding solutions in the Schr\"{o}dinger picture, we would like to draw the reader's attention to one further point. The continuity of $\mathcal{R}$, along with a discontinuity of the scalar field velocity means that the conventionally quantized Mukhanov variable, $v=z\mathcal{R}$ ($z=a\dot{\phi}/H$), is not continuous, and so is an inappropriate choice for quantization in systems with a sudden transition from one background cosmology to another.

%The machinery of the Schr\"{o}dinger picture is somewhat more suited to specification of the initial conditions for the inflationary wavefunction than the Heisenberg or interaction pictures (see for example \cite{Eboli:1988qi, Guven:1998, Fischler:1985ky}). Once these initial conditions are chosen, the evolution of $P_{\cR}(k)$ is given by the corresponding unitary operator of the above Hamiltonian, derived in \cite{BenAryeh:2008bi, Lo:1993}:

\subsection{Unitary Evolution}
Once initial conditions are chosen, the evolution of $P_{\cR}(k)$ is given by the corresponding unitary operator of the Hamiltonian in \ref{hoe}, derived in \cite{BenAryeh:2008bi, Lo:1993}
\begin{equation}
U(t,t_0) = \exp{\left[ c_1(t) J_{+}\right]} \exp{\left[ c_2(t) J_0 \right]} \exp{\left[ c_3(t) J_{-} \right]}\, ,
\end{equation}
where
\begin{align}
J_{+}&=\frac{1}{2} \cR^2\, , & J_0 &=\frac{i}{4}( \cR \Pi_{\cR} + \Pi_{\cR} \cR )\, . & J_{-}&=\frac{1}{2} \Pi_{\cR}^2\, .
\end{align}
The $c_{i}(t)$ can be written as:
\begin{align}\label{ceq}
c_1(t)&=i\,m(t) \frac{\partial}{\partial t } \ln S(t)\, , & c_1(t_0)&=0\, , \nonumber\\
c_2(t)&=-2\ln\frac{S(t)}{S(t_0)}\, , \nonumber\\
c_3(t)&=-i\,S^2(t_0) \int_{t_0}^t \frac{du}{m(u)S^2(u)}\, ,
\end{align}
where $S(t) $ satisfies the differential equations
\begin{align}
\frac{d^2 S(t)}{dt^2}+\xi(t) \frac{d S(t)}{dt}+\omega^2(t)S(t)&=0\, , & \xi(t)&=\frac{\partial}{\partial t}\ln m(t)\, .
\end{align}
Although $S(t)$ will depend on two arbitrary constants, the functions $c_i$ only depend on one fixed by the requirement that $c_1(t_0)=0$.

We will be interested in a cosmology for which one can solve  these equations analytically, it will be described in the next section.  The initial state will be restricted to a gaussian wave functional.
\begin{equation}
\Psi(t)  = N(t) \exp\left[-\frac{1}{2} \int \, \frac{d^3\mathbf{k}}{\left(2\pi\right)^3}\,\cR(\mathbf{k})\, F(k,t)\, \cR(-\mathbf{k})\right]\, .
\end{equation}
Different initial conditions refer to different choices of $F(k, t_0)$. As already anticipated in the introduction and clarified in section \ref{initialcond} there are strong constraints in the form of this function.

The unitary operator defined above evolves a gaussian into a gaussian with different width and normalization \cite{Lo:1993}. The evolution of the width is given by
\begin{align}
\Re F(k,t)&= e^{c_2(t)} \frac{\Re F(k,t_0) }{ 1 + |c_3(t)|^2 |F(k,t_0)|^2 - 2 \, i \, c_3(t) \, \Im F(k,t_0)}\, , \nonumber\\
\Im F(k,t)&=-\frac{c_1(t)}{i}   + e^{c_2(t)} \frac{ \Im F(k,t_0)- i \, c_3(t) \, |F(k,t_0)|^2 }{ 1 + |c_3(t)|^2 |F(k,t_0)|^2 - 2 \, i \, c_3(t)  \,\Im  F(k,t_0)}\, .
\end{align}
And the power spectrum is
\be
P_{\cR}(k,t) = \frac{1}{2 \, \Re F(k,t)}\, .
\ee
For a slow roll cosmology the power spectrum becomes constant soon after the mode $k/{a(t)}$ exits the horizon.

\section{Initial Conditions}\label{initialcond}
As anticipated in the introduction, the initial wave function is not without constraints (see among many others \cite{Eboli:1988qi, Kaloper:2002cs}). Assuming an UV-cutoff $M$ above which classical gravity and quantum field theory are not the operating rules any longer, an effective description implies that at any instant of time the wavefunction has to satisfy the following renormalizability condition: any divergence that appears in $\langle T_{\mu \nu}\rangle$ has to be written in terms of the following linear combination of covariant tensors (for some details see \cite{Birrell:1982ix}):
\begin{equation}
G_{\mu \nu} + \Lambda \, g_{\mu \nu} + g \, H_{\mu\nu} + g_1 \,{}^{(1)} H_{\mu \nu} + g_2 {}^{(2)}H _ {\mu \nu} =  8 \pi G\langle T_{\mu \nu}\rangle\, .
\end{equation}
$G_{\mu \nu}$ is the Einstein tensor and the other tensors, derivatives or higher curvature terms,  are defined in appendix \ref{renorm} . The physical wavenumber $k/a(t)$ redshifts  with time, hence the corresponding initial condition needs only to be fixed once, when this quantity is below M. This procedure differs from regular quantum mechanics in that the Hilbert space grows with time: new wavenumbers are incorporated into the effective evolution as soon as $k/{a(t)}< M$.
For large $k/a$ the function $F(k,t)$ can be expanded as:
\begin{align}
\Re F&= m(t)\omega(t)\left\{1 + \sum_{n\geq 1} \, b_n \, \left( \frac{a H }{k}\right) ^{ n} \right\}\, , \\
\Im F&= m(t)H\left\{\sum_{n\geq 0} \, c_n \, \left( \frac{a H }{k}\right) ^{ n} \right\}\, .
\end{align}
The only constraint for this expansion has been that the wavefunction for $ k/(aH)\gg1$ should look like the flat one. It is clear that this expansion satisfies this constraint  for arbitrary values of the coefficients $b_n$ and $c_n$.  The renormalizability constraint has been explicitly worked out in appendix \ref{renorm}, for a Robertson-Walker cosmology with an equation of state $p=w\rho$ in an $d+1$ dimensional space-time, with a result that turns out to be very restrictive. Indeed, imposing renormalizability along with satisfiability of the Schr\"{o}dinger equation one obtains:
\begin{align}\label{Fconditions}
\Re F&=m(t) \omega(t) \left\{ 1+   \frac{1}{8} (d-1) ( d \, w -1 ) \left( \frac{a H }{k} \right) ^2 \right.\nonumber \\
&-\left.\frac{1}{128} ( d-1) ( d\, w-1) ( -5 + ( d( 3 + 2 w) ) ( -5 + d( 2 + 3 w) ) \left(  \frac{a H }{k} \right) ^4+\cdots \right\}\, , \nonumber\\
\Im F&=m(t) \, H \, \frac{(d-1)}{2} \, \left\{ 1-   \frac{1}{8}  ( d \, w -1 ) ( -2 + d ( 1 + w ) ) \left( \frac{a H }{k} \right) ^2 \right.\nonumber \\
&+\left. \frac{1}{32}  (-1 + d w) (-2 + d + d w) \left(13 - 13 d (1 + w) + d^2 (3 + w (7 + 3 w))\right) \left(  \frac{a H }{k} \right) ^4 + \cdots \right\}\, .
\end{align}

The coefficients not displayed above and represented by the ellipsis are not fixed by the renormalizability constraint, but their values will not, for a reasonable number of e-foldings, affect the form of the power spectrum. Notice that  only  even powers of $k/aH$ appear in the first few terms. Odd powers will produce divergences that cannot be expressed in terms of covariant tensors. As time evolves and the scale factor $a(t)$ grows, physical wavelengths move from being smaller than the  $1/M$ to being bigger than it. Their initial condition is given by the renormalizability constraint in the cosmology, that is for $w$, when they become larger than $1/M$. Clearly, the renormalization doesn't restrict the shape of the wave function for $k/aH\sim 1$ or smaller.

For $d=3$ and $w=-1$, the function above corresponds to:
\begin{align}
\Re F&=m(t)\omega(t) \left\{ 1- \left( \frac{a H }{k} \right) ^2  + \left(  \frac{a H }{k} \right) ^4+\cdots \right\}\, , \nonumber\\
\Im F&=m(t)H \left\{ 1-  \left( \frac{a H }{k} \right) ^2+  \left(  \frac{a H }{k} \right) ^4 + \cdots \right\}\, .
\end{align}
This should be compared with the Bunch-Davis vacuum \cite{Birrell:1982ix}, where:
\begin{align}
\Re F (t)&=m(t)\omega(t)\frac{\left(\frac{k}{aH}\right)^2}{1+\left(\frac{k}{aH}\right)^2}\, , \\
\Im F (t)&=m(t)H\frac{\left(\frac{k}{aH}\right)^2}{1+\left(\frac{k}{aH}\right)^2}\, .
\end{align}
In the limit $k/{aH} \gg 1$ these two expressions coincide. The Bunch-Davis vacuum is a solution of the Schr\"{o}dinger equation that corresponds to  initial conditions with  definite values of the terms represented by the ellipsis.  There is no physical reason to assume any knowledge of the wave function for those values of $k/{aH}$ for which the ellipsis terms are relevant. However, for a large number of e-foldings the value of these coefficients does not make a difference.

For $d=3$ and $w=1$, corresponding to the kinetic energy dominated period  the expansion in equation \ref{Fconditions} reads
\begin{align}\label{abdv}
\Re F&=m(t)\omega(t) \left\{ 1+ \frac{1}{2}\left( \frac{a H }{k} \right) ^2  -\frac{25}{8} \left(  \frac{a H }{k} \right) ^4+\cdots \right\}\, , \nonumber\\
\Im F&=m(t)H \left\{ 1-  \left( \frac{a H }{k} \right) ^2+ 13  \left(  \frac{a H }{k} \right) ^4 + \cdots \right\}\, .
\end{align}
The first few terms of the expansion can be written in compact form as
\begin{equation}
\Re F (t) + i \, \Im F(t) = i \,  m(t) \omega(t) \, \frac{ H_1^{(1)}( x/2) }{H_0^{(1)} (x/2)}\, ,
\end{equation}
where the $H_i(x/2) $ are Hankel functions and $x=k/aH$.

\subsection*{Observability}
Although it is clear from the above that when the inflationary era begins modes which have a physical size less than $1/M$ are essentially in the Bunch-Davies vacuum, this is not enough to guarantee that the initial state of the inflationary wavefunction will not affect observables. After all, modes whose physical size is greater than $1/M$ at the start of inflation do not have constraints from short distance inflationary physics, and for a suitable choice of parameters these may well be the modes that we observe today.

Of course, in an expanding universe we know that modes that were larger than $1/M$ at the start of inflation must have been smaller than that scale at some earlier point. However, it is clear from equation \ref{Fconditions} that constraints on these modes will depend on the nature of the cosmology at that time and will, in general, be different. The observable consequences of such differences are discussed in more detail in section \ref{observables}.

\begin{figure}
\centering
\includegraphics{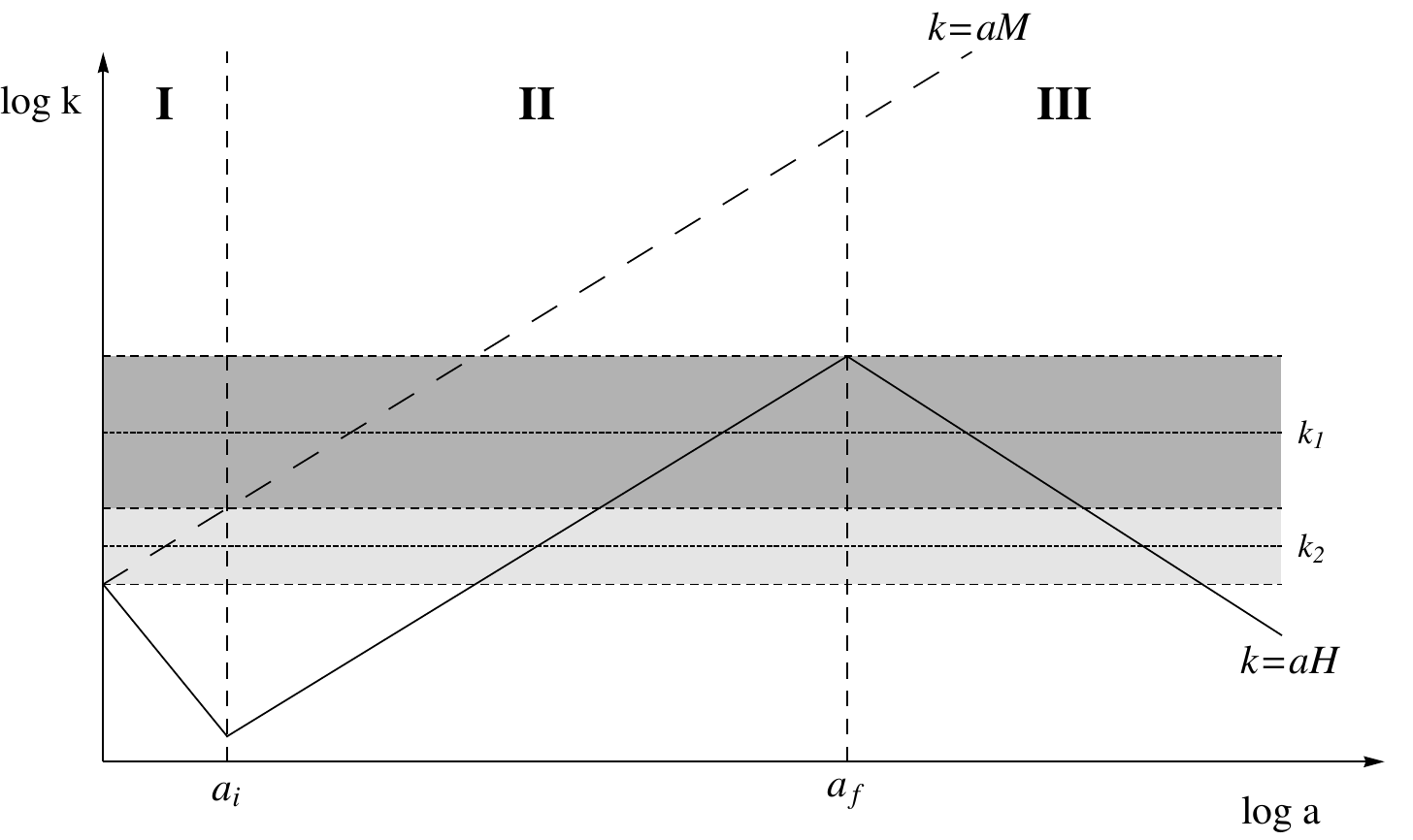}
\caption{A toy universe cosisting of a pre-inflationary, decelerating period (region $I$), inflation (region $II$) and another, post-inflationary, decelerating era (region $III$). The dashed and solid lines show, respectively, the evolution of the comoving cutoff ($aM$) and the comoving Hubble radius ($aH$). Modes above the solid line are within the Hubble radius and accessible to observers. Of these, those that lie in the light grey region have their initial conditions set during a pre-inflationary, decelerating era, whereas those modes in the dark grey region emerged from the cutoff scale during inflation.}
\label{comovingscales}
\end{figure}

In order to more clearly illustrate how different scales have their initial conditions during different epochs, it is helpful to consider the evolution of comoving scales corresponding to the cutoff $aM$, the Hubble radius $aH$ and today's observable modes. In figure \ref{comovingscales} we plot $\log k$ against $\log a$ for the aforementioned scales - region $I$ is a pre-inflationary decelerating epoch, region $II$ corresponds to inflation and region $III$ is the post-inflationary decelerating universe. When modes are above the solid line representing $k=aH$, they lie within the Hubble radius and so are observable (and evolving with time). The initial condition for any particular comoving wavenumber, $k$, is set when its corresponding physical size is equal the the cutoff, $M$. In the plot, this happen when the horizonal line for some given wavenumber intersects the dashed line showing the evolution of $aM$, the comoving cutoff scale. For example, the comoving scale $k_1$ (and any other scale in the dark grey region) has its initial condition set during the inflationary era, while the comoving scale $k_2$ (and others in the light grey region) have their initial states set during the pre-inflationary epoch.

\begin{figure}
\centering
\includegraphics{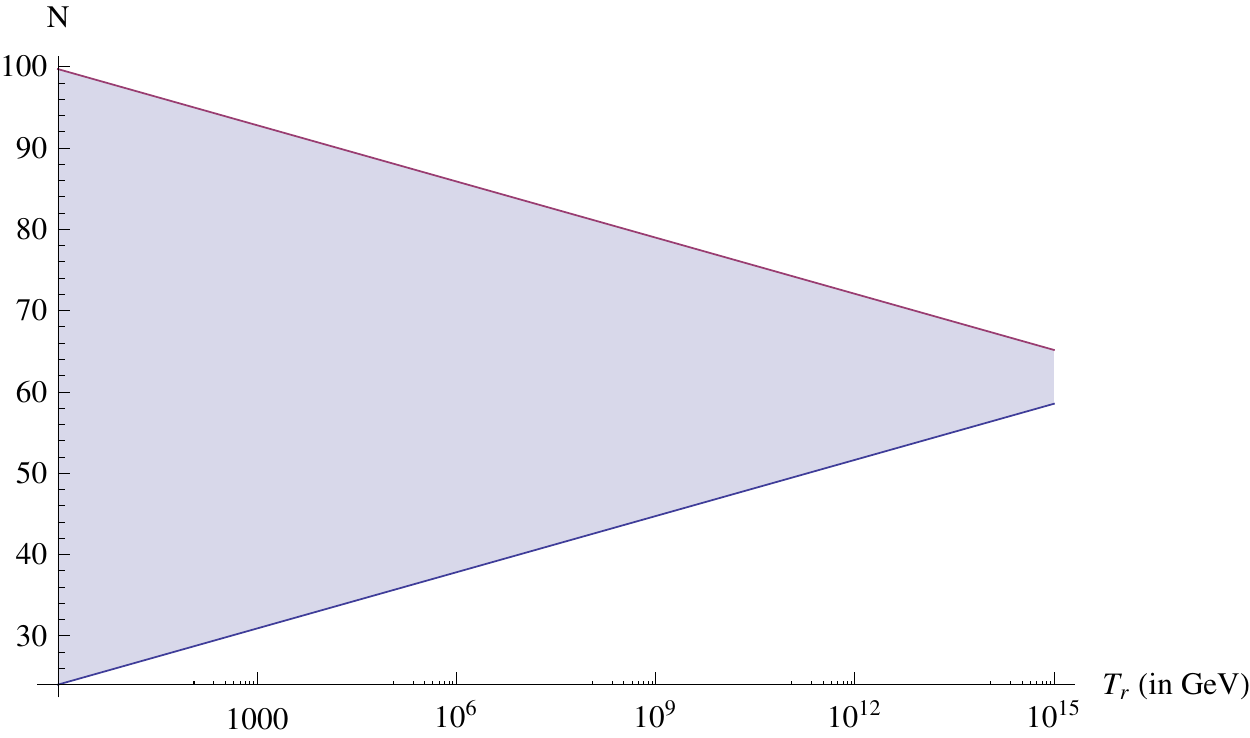}
\caption{Range of e-foldings where initial conditions for observable modes is not necessarily set by UV constraints of inflationary vacuum. The cutoff has been set to the Planck scale, lower cutoffs will shift the upper bound down by an $\mathcal{O}(1)$ amount.}
\label{nboundsfig}
\end{figure}

The range of e-foldings for which observable modes may be larger than $1/M$ at the beginning inflation depends on both $M$ and on the scale of inflation. For clarity, we will assume that reheating at the end of inflation is instantaneous and set the reheat temperature $T_r$ to the inflationary scale. Now, observable scales today have a physical size from (approximately) $10^4\,\textrm{Mpc}$ down to $1\,\textrm{Mpc}$ \cite{Komatsu:2010fb}. This corresponds to $S_1=10^{61}\frac{1}{M}\frac{M}{m_p}$ and $S_2=10^{57}\frac{1}{M}\frac{M}{m_p}$. Requiring that these scales were sub-Planckian at the start of inflation sets a limit on how much the universe can have expanded since that point. Moreover, since we know how much the universe has expanded since the end of inflation (this will, of course, depend on the reheat temperature), it is easy to bound the maximum number of e-foldings as:
\begin{align}
S_2\frac{a_i}{a_{\textrm{today}}}&>\frac{1}{M}\nonumber\\
\Rightarrow S_2\frac{a_i}{a_f}\frac{T_{\textrm{today}}}{T_r}&>\frac{1}{M}\nonumber\\
\Rightarrow e^{N}&<10^{57}\frac{M}{m_p}\frac{T_{\textrm{today}}}{1\,\textrm{GeV}}\frac{1\,\textrm{GeV}}{T_r}\nonumber\\
\Rightarrow N&<102-\log\left[\frac{T_r}{1\,\textrm{GeV}}\right]+\log\left[\frac{M}{m_p}\right]\, .
\end{align}
Subscripts $i$ and $f$ refer to the end of inflation, and ``today'' to the present day. There is also a bound on the minimum number of e-foldings (for a given reheat temperature) that follows from the requirement that there must have been sufficient e-foldings to ensure that inflationary perturbations source the entire observable universe. This is given by:
\begin{align}
N&>\log\left[\frac{T_r}{0.058h\,\textrm{eV}}\right]\nonumber\\
N&>24+\log\left[\frac{T_r}{1\,\textrm{GeV}}\right]
\end{align}
Thus, for a given reheat temperature, the initial conditions for observable modes are not set by UV requirements of the inflationary vacuum if the number of e-foldings is in the following range:
\begin{equation}
24+\log\left[\frac{T_r}{1\,\textrm{GeV}}\right]<N<102-\log\left[\frac{T_r}{1\,\textrm{GeV}}\right]+\log\left[\frac{M}{m_p}\right] \, .
\end{equation}
These bounds are illustrated in figure \ref{nboundsfig} for a Planck scale cutoff (lowering the cutoff scale will have an $\mathcal{O}(1)$ effect).

That the initial form of the wavefunction for observable modes may not necessarily be of the Bunch-Davies form is, in and of itself, interesting. It is, however, illuminating to construct an explicit model where we can see how a prior epoch can set these initial conditions. We discuss such a model and its observable consequences in the following sections.

\section{The Model and its Evolution}\label{modelevol}
\subsection{Cosmology}
Because we are interested in a model with an analytical solution, we will consider the following potential:
\begin{equation}\label{cos}
V(\phi)=
\left\{
\begin{array}{ccl}
0 & \phi< \phi_{i}  & \textrm{region } I   \\
\Lambda^4\left(1-\alpha\,\phi\right)\  & \phi_{i} < \phi < \phi_{f}  & \textrm{region } II\, ,    \\
0  & \phi>\phi_{f} & \textrm{region } III
\end{array}
\right.
\end{equation}
with the assumption that  $\alpha\phi\ll 1$ and the initial condition on $\phi$ is such that its kinetic energy is of order $\Lambda^4$ when the inflaton field reaches the value $\phi_i$. This condition requires  fine-tuning of the initial condition on $\dot{\phi}$. With these assumptions
\begin{equation}
\frac{1}{2} \dot{\phi}^2+ V(\phi) \sim
\left\{
\begin{array}{ccl}
\frac{1}{2} \dot{\phi}^2 & \phi< \phi_{i}    \\
 \Lambda^4\left(1 - \alpha\,\phi\right)\  & \phi_{i} < \phi < \phi_{f}\, .  \\
\frac{1}{2} \dot{\phi}^2 & \phi>\phi_{f}
\end{array}
\right.
\end{equation}
In terms of the FRW time, $t$, the time dependence of the scale factor is
\begin{equation}
a(t)=
\left\{
\begin{array}{ccc}
a_0\left(1+3H\left(\frac{a_i}{a_0}\right)^3\left(t-t_0\right)\right)^{1/3}  & t_0\leq t \leq t_i\\
a_i e^{H(t-t_i)}  & t_i\leq t \leq t_f\, .\\
a_i e^N \left(1+3H(t-t_f)\right)^{1/3}  & t_f \leq t
\end{array}
\right.
\end{equation}
$a_0$ is the scale factor when the initial condition is set, $\abi$ is the scale factor at the beginning of inflation, $N$ is the number of e-foldings, $H$ is the value of the Hubble constant during inflation ($H^2=\Lambda^4/3$ -- in units of $m_{P}^{\textrm{reduced}}=1$), $t_0$ is the time when the initial condition is set and $t_i$($t_f$) are the times when inflationary period starts(ends)\footnote{$a_0$, $a_i$, $t_i$ and $t_0$ are not independent parameters: $t_i-t_0=\frac{1}{3 H } \left[1-\left(\frac{a_0}{a_i}\right)^3\right]$}. We assume a sudden transition between the different periods.

The equation of motion for the inflaton can be written as
\begin{equation}
\ddot{\phi} + 3 \, H(t) \, \dot{\phi} + V'(\phi) =0\, .
\end{equation}
In regions $I$ and $III$ this implies that
\begin{equation}
\frac{d}{dt} ( a^3 \dot{\phi}) = 0\, ,
\end{equation}
while in region $II$, we shall assume slow-roll
\begin{equation}
\dot{\phi} = H\alpha\, .
\end{equation}
Just after the transition, $\dot{\phi}$ depends on time, but stabilizes to this value in few e-foldings.
\subsection{Evolution}
As mentioned previously, we concern ourselves solely with Gaussian wavefunctions, with the initial condition:
\begin{equation}\label{initialwavefunction}
\psi_{\mathbf{k}}(t_i)=N_{\mathbf{k}}(t_i)\exp{\left[-\frac{1}{2}\cR({\bf k})F_B\cR(-{\bf k}) \right]}\, .
\end{equation}
Since, at this order, the different $k$ are decoupled, we have displayed only the time evolution of a fixed value of $k$; we shall continue in this vein for clarity.

By using the techniques outlined in section \ref{schroedpicture} and solving the differential equations given in equation \ref{ceq} it is possible to calculate the evolution of the wavefunction in equation \ref{initialwavefunction} for the model outlined above. The explicit forms of the $c_i$'s are given in appendix \ref{cevol} and the results for the wavefunction evolutions in the three regions of our cosmology are below.

\subsubsection{Evolution during the period before inflation starts}
With the above initial form, the wave function at the beginning of the inflationary period, when $\phi$ reaches the value $\phi_i$, will be
\begin{equation}\label{ga}
\psi_{\mathbf{k}}(t_i)=N_{\mathbf{k}}(t_i)\exp{\left[-\frac{1}{2}\cR({\bf k})F_1(k,t_i)\cR(-{\bf k}) \right]}\, .
\end{equation}
The real and imaginary parts of $F_1$ are given by
\begin{align*}
\Re F_1(t_{i})&=\frac{w_0^2\lambda^2(\Re F_0)G_2(\wbi,w_0)^2}{H^2\abi^6(\Re F_0)^2G_1(\wbi,w_0)^2+\left(w_0\lambda G_2(\wbi,w_0)-\abi^3H(\Im F_0)G_1(\wbi,w_0)\right)^2}\, ,\\
\Im F_1(t_{i})&=\frac{\wbi G_3(\wbi,w_0)}{\abi^3H G_2(\wbi,w_0)}
+\frac{\left(w_0^2\lambda^2(\Im F_0)-\abi^3H\left((\Re F_0)^2+(\Im F_0)^2\right)\frac{G_1(\wbi,w_0)}{G_2(\wbi,w_0)}\right)G_2(w_0,w_0)^2}{H^2\abi^6(\Re F_0)^2G_1(\wbi,w_0)^2+\left(w_0\lambda G_2(\wbi,w_0)-\abi^3H(\Im F_0)G_1(\wbi,w_0)\right)^2}\, .
\end{align*}
With
\begin{align*}
\lambda&= a_0^6\dot{\phi_0}^2\, , & w_0&= \frac{k}{a_0H_0}\, , & \wbi&= \frac{k}{a_iH_i}\, , & \Re F_1(k,t_0)&=\Re F_0\, , & \Im F_1(k,t_0)&=\Im F_0\, ,
\end{align*}
and
\begin{align*}
G_1(\wbi,w_0)&= J_0 ( \frac{w_0}{2}) Y_0 ( \frac{\wbi}{2}) -  J_0 ( \frac{\wbi}{2}) Y_0 ( \frac{w_0}{2})\, , \\
G_2(\wbi,w_0)&= J_1 ( \frac{w_0}{2}) Y_0 ( \frac{\wbi}{2}) -  J_0 ( \frac{\wbi}{2}) Y_1 ( \frac{w_0}{2})\, , \\
G_3(\wbi,w_0)&= J_1 ( \frac{w_0}{2}) Y_1 ( \frac{\wbi}{2}) -  J_1 ( \frac{\wbi}{2}) Y_1 ( \frac{w_0}{2})\, .
\end{align*}
In region $I$, $H(\e)=\frac{1}{6}\dot{\phi}^2$ and $a^3\dot{\phi}$ is conserved, so the continuity of $a(t)$ and $H(t)$ at $t_i$ implies:
\begin{equation}
\lambda= 6a_0^6H_0^2=6\abi^6H^2\, .
\end{equation}
For a given $k$, if $w_0 \gg 1$ then $\wbi \gg1$. In this limit
\begin{align*}
G_1( \wbi, w_0)&\rightarrow\frac{4}{\pi} \frac{1}{\sqrt{\wbi w_0}} \sin{(\wbi-w_0)}\, , \\
G_2( \wbi, w_0)&\rightarrow\frac{4}{\pi} \frac{1}{\sqrt{\wbi w_0}} \cos{(\wbi-w_0)}\, , \\
G_3( \wbi, w_0)&\rightarrow-\frac{4}{\pi} \frac{1}{\sqrt{\wbi w_0}} \sin{(\wbi-w_0)}\, ,
\end{align*}
and
\begin{align}\label{ichm}
\Re F_1(t_{i}) & =6a^2_{i}k\, , \\
\Im  F_1(t_i)  & = 6 a^3_i H\, .
\end{align}
This is the expected result since the evolution from $t_0$ to $t_i$ is adiabatic, so the wavefunction at $t_i$  should be the ground state of the harmonic oscillator (equation \ref{hoe}) at time $t_i$.

There are values of $k$ for which $w_0\sim\mathcal{O}(1)$, while $\wbi\gg1$. For these values no prediction can be made. Because $w_i \gg 1$ its behavior should be insensitive to the curvature of the space-time, but because $k/{a_i} \ll M$ we can't expect the renormalization constraint to select a particular behavior.

\subsubsection{Evolution during inflation}
During the inflationary period the wave function stays gaussian
\begin{equation}
\psi_{\mathbf{k}}(t)=N_{\mathbf{k}}(t)\exp{\left[-\frac{1}{2}\cR({\bf k})F_{2}(k,t)\cR(-{\bf k})\right]}\, .
\end{equation}
The real part of $F_{2}$ is given by:
\begin{equation}\label{eveq}
\Re F_{2}(t) = \frac{y_{i}^2\Re F_1(t_{i})}{\xi^2 ( ReF_1(t_i)^2 + Im F_1(t_i)^2) G_4(y,y_{i})+ G_5(y,y_{i})- 2 \xi \sqrt{G_4(y,y_i) G_5(y,y_i)} Im F_1(t_i)}\, . \end{equation}
With
\begin{align*}
\xi&=\frac{9 \, H^3 }{\abi^3\alpha^2 y^2_{i}}\, , & y_{i}&=\frac{k}{a_{2}(t_{i})H}=w_i\, , & y&=\frac{k}{a_{2}(t)H}\, ,
\end{align*}
and
\begin{align*}
G_4(y,y_{i})&= \left[(y-y_{i})\cos(y-y_{i}) - (1 + y y_{i})\sin(y-y_{i})\,\right]^2\, , \\
G_5(y,y_{i})&= \left[y \cos(y-y_{i}) - \sin(y-y_{i})\right]^2\, .\\
\end{align*}

In writing equation \ref{eveq} we have assumed a sudden transition between the kinetic energy dominated era and the inflationary era. Let the actual transition time be $\tau$. The sudden approximation is  suitable  for frequencies that are much smaller than $1/\tau$. For frequencies larger than $1/\tau$ the transition is adiabatic. To say which regime the observed wavelengths correspond requires a knowledge of the number of e-foldings $N$, the reheating temperature $T_r$, and the value of $\tau$.  Figure \ref{nboundsfig} makes it clear that there is a reasonable region of parameter space where the sudden approximation can make sense.

When a sudden transition occurs (and we use equation \ref{ichm}), the $k$ dependence of $\Re F_{2}(t)$ is neither proportional to $k$ nor to $k^3$. These behaviors emerge at very specific times. For $t =t_{i}$, $\Re F_{2}(t_i)=\Re F_{1}(t_{i})$ and hence for those values of $k$ for which $\Re F_1\sim k$ so is $\Re F_{2}$. For very late times, modes that are well outside the horizon $y\ll1$ but were  well inside the horizon at the beginning of inflation, $y_{i}\gg1$, have
\begin{align*}
G_4(y,y_{i})&\sim   y_{i}^2\cos^2y_{i}\ , \\
G_5(y,y_{i})&\sim  \sin^2y_i\, .
\end{align*}
So,
\begin{equation}\label{k3d}
\Re F_{2}(t) \sim \frac{k^3}{6H^2\left(\frac{H}{\dot{\phi}}\right)^4 \cos^2(\wbi)}\, .
\end{equation}
Notice the following:
\begin{itemize}
\item The dominant behavior is $k^3$, even when the initial state is not the BD state as long as the short wavelength behavior is fixed by the renormalization constraint.  For those modes, mentioned at the end of the last section, for which $w_0 \sim O(1), w_i \gg1$ this $k^3$ dependence is not guaranteed and it depends on the initial conditions, for these momenta the renormalization constraint does not apply.
\item The power spectrum  is modulated by the presence of $\cos^2(\wbi)$. The frequency of this oscillation is extremely high (it is the inverse of the comoving horizon size at the start of inflation and thus many orders of magnitude larger than the highest observed wavenumber) thus current experiments can't measure the dependence in $k$, only its average valued over an interval of k. It is the sharp variation of the mass in the transition form kinetic energy domination to inflation that is responsible for the appearance of this term.
\item The normalization is different, the usual Bunch-Davis vacuum  will include a factor of $\left(\frac{\dot{\phi}}{H} \right)^2$ instead of the factor $\left(\frac{\dot{\phi}}{H}\right)^4$ in equation \ref{k3d}. The other difference is the appearance of the factor $6$. These two effects are tied to each other and to the sudden approximation. $\frac{H}{\dot{\phi}}$ changes very fast when going from kinetic domination to inflation, its value jumping from $1/\sqrt{6}$ to  $\frac{H}{\dot{\phi}}$\footnote{We thank R. Flauger for pointing out that a similar effect appears in Axion Monodromy Inflation \cite{Flauger:2010ja}.}.
\item As in standard inflationary scenarios, the value of $\Re F_2$ does not change as long as the modes are outside the horizon, $y \ll 1$.
\end{itemize}
There are instances when the  inflation period does not erase completely the memory of the initial conditions. Indeed, the first two items above reflect a dependence on the initial conditions.

We can summarize the differences between sudden and adiabatic approaches to inflation with the following expression for the power spectrum:
\begin{equation}\label{fr}
P_{\cR}(k) = c(k) \, \left( \frac{H}{\dot{\phi}} \right)^{\beta(k)} H^2 \, \frac{1}{k^3}\, .
\end{equation}
The coefficients $\beta$ and $c$ are as follows:
\begin{itemize}
\item $\beta=4, c=3$ if the sudden approximation is applicable : $\frac{k}{a(t_i) } \ll \frac{1}{\tau}$.
\item $\beta=2, c=1/2$ when the sudden approximation is not applicable $\frac{k}{a(t_i) } \gg \frac{1}{\tau}$.
\end{itemize}
The above formulae readily give the following expression for the spectral index $n_s$:
\begin{equation}\label{nscalar}
n_s-1=\beta \eta - ( 2 +\beta) \epsilon \, .
\end{equation}

A numerical analysis is required to determine the behavior of $\beta(k)$  for the interpolating values between these two regimes, with a continuous model for the transition between decelerating and accelerating expansion which can encompass both sudden and adiabatic behaviour. While such an analysis is not the focus of this paper, we have carried out a condensed numerical study in order to confirm the that assumptions made above are consistent and reasonable. This analysis, which shows the existence of distinct of sudden and adiabatic regimes, can be found in appendix \ref{numericalresults}.

\subsection{Tensor Modes}
Before moving on to a discussion of the observability (or otherwise) of the initial condition effects discussed above, let us briefly consider tensor modes in these models. The second order action for the tensor modes obtained from expanding the Einstein-Hilbert action is \cite{Baumann:2009ds}:
\begin{equation}
S= \frac{1}{8} \int \, dt \, d^3x \, a^3(t) \left\{ ( \dot{h_{ij}})^2 - \frac{1}{a^2}( \partial_k h_{ij})^2 \right\}\, .
\end{equation}
The evolution of these modes is derived from the Hamiltonian
\begin{equation}
\mathcal H = \frac{1}{2 m(t)} \,\, \Pi_{h}^2 + \frac{1}{2} m(t) \,\, \omega^2(t) h^2\, ,
\end{equation}
where
\begin{equation}\label{mfd}
m(t) \equiv \frac{a^3(t)}{4}  \hspace{4ex} \mbox{and} \hspace{4ex} \omega(t) = \frac{k}{a(t)}\, .
\end{equation}
Notice that this action is independent of $\dot{\phi}$, thus the power spectrum (and its spectral index, $n_t$) will be unchanged from the standard result:
\begin{align}
P_t(k)&=4\frac{H_*^2}{k^3}\, ,\label{ptensor}\\
n_t&=-2\epsilon\, . \label{ntensor}
\end{align}
The subscript $*$ denotes the value at horizon crossing. Combining this with the expression derived above for the scalar power spectrum gives the scalar-tensor ratio:
\begin{equation}\label{rtensorscalar}
r=\frac{4\left(2\epsilon\right)^{\frac{\beta}{2}}}{c}\, .
\end{equation}

\section{Observables}\label{observables}
\begin{figure}
\centering
\includegraphics[width=\textwidth]{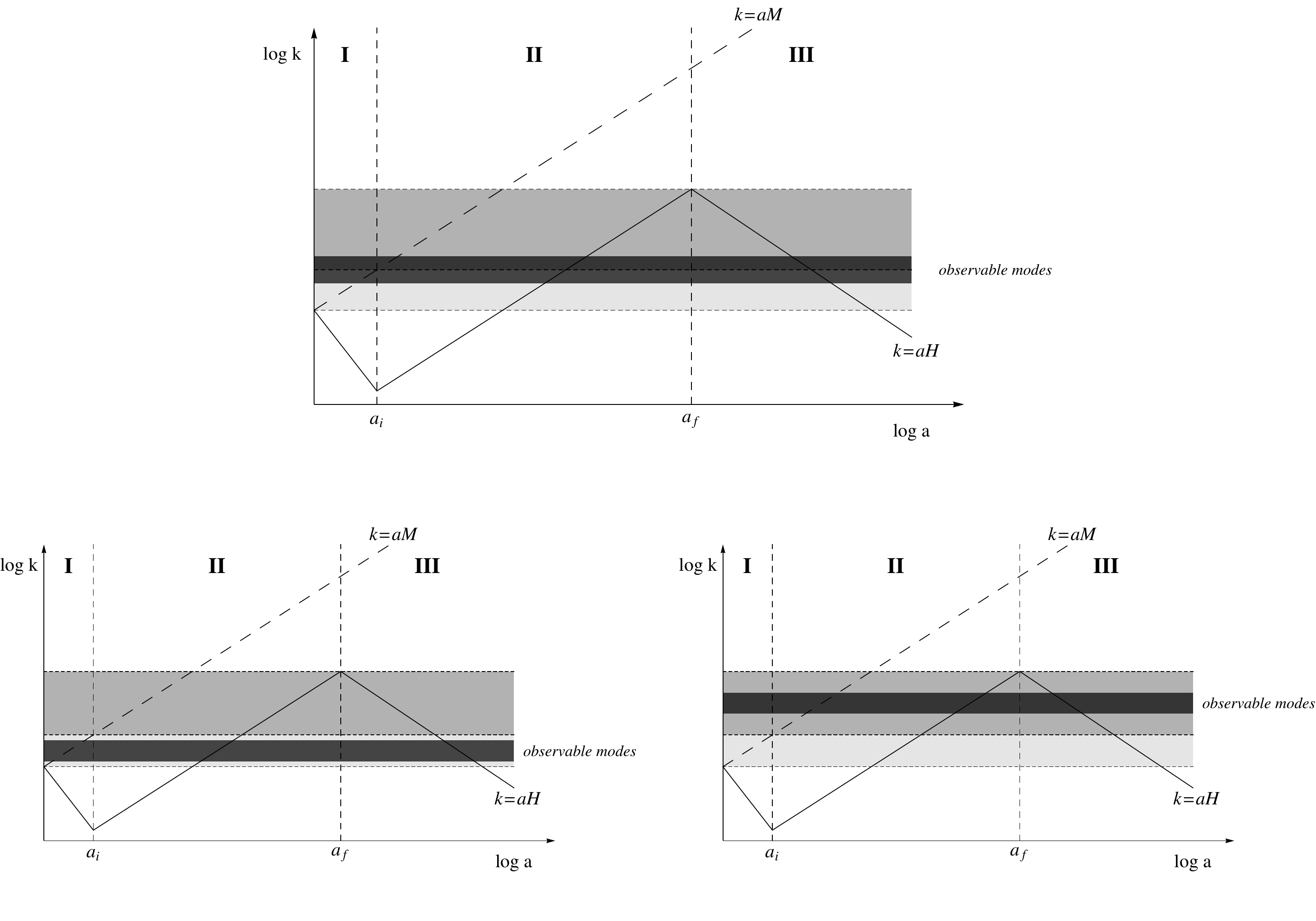}
\caption{As in figure \ref{comovingscales}, these plots illustrate the evolution of various relevant comoving scales for our model. The light grey denotes modes that emerge from the cutoff scale before inflation, the medium grey those modes that emerge during inflation and the dark grey those modes that are observed. At the bottom, on the left the observable modes have their initial conditions set during the pre-inflationary epoch, and on the right initial conditions are set during inflation. The top plot illustrates the most dramatic scenario, where both possibilities are visible.}
\label{comovingscales2}
\end{figure}

We have found that if a pre-inflationary epoch of kinetic domination sets the initial conditions for some subset of observable modes, these modes will have a different power spectrum. The clearest difference is the amplitude of the power spectrum, where instead of a $H^2/\epsilon$ dependence, there is a $H^2/\epsilon^2$ dependance. If this model reflects reality, there are three possible scenarios we should consider: observable modes have their initial conditions set before inflation, during inflation or a combination of both. These scenarios are illustrated in figure \ref{comovingscales2}.

The most dramatic signature would occur if a fraction of the observable modes have their UV behaviour (and thus their amplitudes) set before inflation, and the rest during inflation. This is the case shown in the top panel of figure \ref{comovingscales2}. Here we would see a dramatic change in amplitude, with modes having a sudden drop in amplitude (by factor of $1/\epsilon$ below a wavelength set by the comoving size of the cutoff scale at the start of inflation. Such a change would result in sharp features in both the power spectrum amplitude and tilt.

In the case of the tilt, equation \ref{nscalar} implies the change in $n_s-1$ would be $(n_s-1)+2\epsilon$. Such a change is tightly constrained for the observable range of $k$. WMAP7 \cite{Komatsu:2010fb} gives $1-n_s=0.037\pm0.014$, which would rule out a change in $n_s$ of the order predicted. Note, however, a more complete reconstruction of $n_s$ could allow additional flexibility \cite{Peiris:2009wp}, though such flexibility is limited in the case (as we have) of a power law prior.

If observable modes have their UV behaviour set either entirely before or entirely during inflation (the bottom half of figure \ref{comovingscales2}, then distinguishing between the two scenarios is still possible, though not with current data. This can be done by comparing different relationships between various spectral parameters in the two scenarios.

In particular, consider the formulae for the ratio of the scalar to tensor power spectra and their spectral indices:
\begin{align}
n_s-1&=\beta \eta - ( 2 +\beta) \epsilon\, ,\tag{\ref{nscalar}}\\
n_t&=-2\epsilon\, ,\tag{\ref{ntensor}}\\
r&=\frac{4\left(2\epsilon\right)^{\frac{\beta}{2}}}{c}\, .\tag{\ref{rtensorscalar}}
\end{align}
From the scalar spectral index alone it would be difficult to discriminate between the different scenarios. However, if tensor modes are detected along with their spectral index, one finds different consistency conditions relating $n_t$ and $r$. When modes have their initial conditions set by UV inflationary physics, one has:
\begin{equation}
r=-8n_t\, .
\end{equation}
Whereas, if pre-inflationary UV physics sets the initial conditions, then:
\begin{equation}
r=\frac{4}{3}n_t^2\, .
\end{equation}
While the distinction between the above relationships is dramatic, it would be very hard to make such a check if the latter scenario held true. If we take the maximum possible value of $\epsilon$ as $\mathcal{O}(10^{-2})$, then we would have to be able to measure $r$ at values around $\mathcal{O}(10^{-4})$, well beyond the scope of current or planned experiments. Of course, this also means that a detection of tensor modes at level greater than $\sim10^{-4}$ probably rules out the setting of initial conditions by pre-inflationary physics.

\section{Conclusions}\label{conclusions}
We have discussed the possibility of observing the effect of initial conditions for the inflationary wavefunction. As has been previously studied, requiring a renormalizable energy-momentum tensor constrains the form of the inflationary vacuum for modes shorter than some cutoff ($<1/M$) at the beginning of inflation. However, modes larger than this cutoff can have different initial conditions, set during some pre-inflationary era.

To examine this possibility we have considered a particular realization of the above scenario, where observable modes have their initial conditions set by requiring renormalizability of the energy momentum tensor during a decelerating epoch prior to the onset of inflation. In this model, the change in initial conditions has sharp consequences for the spectrum of primordial fluctuations and thus for the observable spectral properties of cosmological perturbations. We find that the scalar spectrum is boosted by an extra factor of $1/\epsilon$ as well as having a different numerical normalization. We discuss in section \ref{observables} that while such difference would be hard to measure from the scalar spectrum, it could be seen if tensor modes are detected.

Although we have worked in a specific model, it is clear that there are three criteria that must be met in order for the effects of a pre-inflationary epoch on the initial conditions for the inflationary wavefunction to be observed.
\begin{enumerate}
\item Inflation must be preceded by an era of decelerating expansion.
\item The transition into inflation must be sudden ($k/a_i\ll\tau$ for observationally relevant modes).
\item There must be sufficiently few e-foldings so the observationally relevant modes have their initial conditions set by the pre-inflationary epoch (as given by the bounds in figure \ref{nboundsfig}.
\end{enumerate}
The last of these criteria is equivalent to the assertion that the Bunch Davies state is an attractor at large numbers of initial conditions. This is consistent with existing work on this subject, but we note that the typical number of e-foldings required to reach the attractor solution can be larger than the minimum number needed to give scale invariance over an observationally suitable range of wavenumbers.

While none of the above are particularly generic, neither do they involve any unreasonable fine tuning -- the numerical example we consider in the appendix has of order $10^{-5}$ tuning in initial conditions, which is large but not by the standards of inflationary model building. As such, it is notable that the much vaunted ability of inflation to erase that which came before can be circumvented to a small but measurable degree in a large and simple class of inflationary models.

\subsection*{Acknowledgments}
The authors would like thank to thank Eiichiro Komatsu for helpful discussions. S. Paban would also like to thank the Aspen Center for Physics for hospitality while part of this work was done. This material is based upon work supported by the National Science Foundation under Grant No. PHY-0969020 and by the Texas Cosmology Center, which is supported by the College of Natural Sciences and the Department of Astronomy at the University of Texas at Austin and the McDonald Observatory.

\appendix
\section{Renormalization}\label{renorm}
The initial conditions for the wave function are very constrained by the requirement that, at any time, the expectation value of the energy momentum tensor, $\langle T_{\mu\nu}\rangle$,  has to be renormalizable \cite{Eboli:1988qi}. This means that the infinities that appear in the computation of $\langle T_{\mu\nu}\rangle$ in $n=d+1$ dimensions, can be subtracted by a renormalization of the constants that appear in the ``generalized" Einstein equation
\begin{equation}
G_{\mu\nu} + \Lambda g_{\mu\nu} + gH_{\mu\nu} + g_1H_{\mu\nu}^{(1)}+g_2H_{\mu\nu}^{(2)} =  8\pi G\langle T_{\mu \nu}\rangle\, .
\end{equation}
After renormalization this reads
\begin{equation}
G_{\mu \nu} = 8\pi G_R\langle T_{\mu \nu}\rangle_R\, .
\end{equation}
Here $G_{\mu \nu}$ is the Einstein tensor and
\begin{align*}
H_{\mu\nu}&= \frac{1}{\sqrt{-g}} \frac{\delta}{\delta g^{\mu\nu}}\int d^n x \sqrt{-g}R^{\alpha\beta\gamma\delta}R_{\alpha\beta\gamma\delta}\, , \\
H_{\mu\nu}^{(1)}&= \frac{1}{\sqrt{-g}} \frac{\delta}{\delta g^{\mu\nu}}\int d^n x \sqrt{-g}R^2\, , \\
H_{\mu\nu}^{(2)}&= \frac{1}{\sqrt{-g}} \frac{\delta}{\delta g^{\mu\nu}}\int d^n x \sqrt{-g}R^{\alpha\beta}R_{\alpha\beta}\, .
\end{align*}
Writing the wave function as in equation \ref{ga} and imposing that it satisfies the Schr\"{o}dinger equation one obtains the following:
\begin{align}
\Re F&=m(t) \omega(t) \left\{ 1+   \frac{1}{8} (d-1) ( d \, w -1 ) \left( \frac{a H }{k} \right) ^2 \right.\nonumber \\
&-\left.\frac{1}{128} ( d-1) ( d\, w-1) ( -5 + ( d( 3 + 2 w) ) ( -5 + d( 2 + 3 w) ) \left(  \frac{a H }{k} \right) ^4+\cdots \right\}\, . \nonumber\\
\Im F&=m(t) \, H \, \frac{(d-1)}{2} \, \left\{ 1-   \frac{1}{8}  ( d \, w -1 ) ( -2 + d ( 1 + w ) ) \left( \frac{a H }{k} \right) ^2 \right.\nonumber \\
&+\left. \frac{1}{32}  (-1 + d w) (-2 + d + d w) \left(13 - 13 d (1 + w) + d^2 (3 + w (7 + 3 w))\right) \left(  \frac{a H }{k} \right) ^4 + \cdots \right\}\, .\tag{\ref{Fconditions}}
\end{align}
With this choice of wave function the expectation value of the energy momentum tensor is:
\begin{align}
\langle T_{00}\rangle &= \frac{1}{2a^{d+1}}g_{00}\int\frac{d^dk}{(2\pi)^d}k +\frac{1}{2a^{d-1}}\left(\frac{d-1}{4d}\right)G_{00}\int\frac{d^dk}{(2\pi)^d}\frac{1}{k} \nonumber \\
&+\frac{1}{8a^{d-3}}\left(\frac{d-1}{4d}\right)^2H_{00}^{(1)}\int\frac{d^dk}{(2\pi)^d}\frac{1}{k^3} \nonumber \\
&-\frac{1}{64a^{d-3}}(d-1)^2(dw-1)(d-3)H^4\int\frac{d^dk}{(2\pi)^d}\frac{1}{k^3}\, , \\
\langle T_{ij}\rangle &= \frac{1}{2a^{d+1}}g_{ij}\left(-\frac{1}{d}\right)\int\frac{d^dk}{(2\pi)^d}k-\frac{1}{2a^{d-1}}\left(\frac{d-1}{4d}\right)G_{ij}\int\frac{d^dk}{(2\pi)^d}\frac{1}{k} \nonumber \\
&-g_{ij}\frac{1}{16da^{d-1}}(d-1)^3H^2\int\frac{d^dk}{(2\pi)^d}\frac{1}{k} \nonumber \\
&+\frac{1}{8a^{d-3}}\left(\frac{d-1}{4d}\right)^2H_{ij}^{(1)}\int\frac{d^dk}{(2\pi)^d}\frac{1}{k^3} \nonumber \\
&-\frac{3}{256a^{d-3}}\frac{(d-1)^2}{d}(dw-1)(d-3)(-3+d(2+w))(1+2w)H^4\int\frac{d^dk}{(2\pi)^d}\frac{1}{k^3}\, .
\end{align}
Thus giving a perfectly renormalizable energy-momentum tensor.

\section{Time evolution of the $c_i$}\label{cevol}
In this section we will give the explicit expressions for the functions $c_i(\eta)$ that were introduced in equation \ref{ceq}.
\begin{itemize}
\item Region I
\begin{align*}
c_1(t)&= iw\left(\frac{a^6_0\dot{\phi}^2_0}{a_i^3H}\right)\frac{J_1(w/2)Y_1(\wi/2)-J_1(\wi/2)Y_1(w/2)}{J_1(\wi/2)Y_0(w/2)-J_0(w/2)Y_1(\wi/2)}\, ,  \\
c_2(t)&= -2\log\left[\frac{J_1(\wi/2)Y_0(w/2)-J_0(w/2) Y_1(\wi/2) }{J_1(\wi/2) Y_0(\wi/2)-J_0(\wi/2) Y_1(\wi/2)}\right]\, , \\
c_3(t)&= -\frac{i}{k}\left(\frac{a_i}{a_0}\right)^2\left(\frac{\abi^4H^2}{a^6_0\dot{\phi}^2_0}\right)\frac{J_0(\wi/2)Y_0(w/2)-J_0(w/2)Y_0(\wi/2)}{J_1(\wi/2)Y_0(w/2)-J_0(w/2)Y_1(\wi/2)}\, .
\end{align*}
\item Region II
\begin{align*}
c_1(t) &= i\frac{k^3\alpha^2\sin(w-\wbi)}{9H^6w\left(w\cos(w-\wbi)-\sin(w-\wbi)\right) }\, , \\
c_2(t) &= -2\log\left[\frac{w\cos(w-\wbi)-\sin(w-\wbi)}{\wbi}\right]\, ,  \\
c_3(t) &= -i\frac{9H^6\wbi^2}{k^3\alpha^2}\left(\frac{1}{\wbi}-\frac{\cos(w-\wbi)+w\sin(w-\wbi)}{w\cos(w-\wbi)-\sin(w-\wbi)}\right)\, .
\end{align*}
\end{itemize}

\section{Matching Conditions in the Heisenberg Picture}

The Israel matching conditions worked out by \cite{Deruelle:1995kd}, in the cosmology context,  will be applied in this section to the approximation, made in this work, of a sharp transition between a kinetic dominated period and an inflationary one.

In a Friedmann-Roberston-Walker  background described by the metric

$$ ds^2 = a^2(\eta) [ -   d\eta^2  +  \delta_{ij} dx^i dx^j ] $$ with perturbations

$$ ds^2 = a^{2}(\eta) [ - ( 1 + 2 \Phi)  d\eta^2 + 2  B_i d\eta dx^i  + ((1- 2 \Psi)  \delta_{ij} + E_{ij})dx^i dx^j ] $$ where

$$ B_i \equiv \partial_i B - S_i, \hspace{4ex} \mbox{such that} \hspace{3ex} \partial^i S_i = 0 $$ and

$$ E_{ij} \equiv 2 \partial_{ij} E + 2 \partial _{(i}F_{j)} + h_{ij},  \hspace{4ex} \mbox{such that} \hspace{3ex} \partial^i F_i=0, h^i_i = \partial ^ih_{ij}=0 $$
Deruelle and Mukhanov \cite{Deruelle:1995kd} proved that Israel matching conditions at a surface $\Sigma:$ $\phi_0(\eta,x^i)+ \delta \phi(\eta,x^i)=\mbox{constant}$, in the Newtonian gauge ($B=0$, $E=0$),  become

\begin{eqnarray}
[ \cR ]_{\pm} &=& 0  \nonumber\\
\left[ \delta \phi / {\phi'_0} \right]_ { \pm} & = & 0   \label{IMC} \\
\left[ \cH \Phi + \Psi' + ( \cH' - \cH^2 ) \delta \phi / {\phi'_0} \right]_{\pm} & = & 0 \nonumber
\end{eqnarray}
where $\left[ A \right]_{\pm} \equiv A_+ - A_-$, is the difference in the variable $A$ across the surface.

In the case at hand the two regimes that are matched across the surface $\phi= \phi_i$ are a kinetic dominated period  and a nearly de Sitter expansion. In this scenario the scale factor evolves as:

\begin{eqnarray*}
a[\eta] &=&  \sqrt{1 + 2 H \eta}  \hspace{4ex}  \eta \leq 0   \\ \\
a[\eta]  &= & \frac{1}{1- H \eta}  \hspace{7ex}   \eta \geq 0
\end{eqnarray*}
The conformal time has been shifted, so that the transition between the two phases happens at $\eta=0$, and the scale factor has been normalized to be equal to 1 at this transition.

 Satifying the classical Israel conditions requires that both the scale factor $a(\eta)$ and $\cH(\eta)=a'(\eta)/{a(\eta)}$ be continuous across the surface $\Sigma$. $H$ is the value of $a \cH$ in the nearly de Sitter phase studied in this work and is assumed constant.
Only two of the conditions (\ref{IMC}) are independent. Indeed, in the Newtonian gauge $\Psi=\Phi$, and  the equations of motion:

\begin{eqnarray*}
\Phi' + \cH \Phi &=& \frac{1}{2} \phi'_0 \delta \phi \\
\cH^2 - \cH' &=& \frac{1}{2} { \phi'_0}^2 \\
%\Delta \Phi & = & -\frac{1}{2} \frac{{ \phi'_0}^2}{\cH} \left( \frac{v}{z} \right)' \\
%\cR &=& \Phi + \frac{\cH}{ \phi'_0} \delta \phi
\end{eqnarray*}
guarantee that the last of the conditions (\ref{IMC})  is trivially satisfied. It is convenient to use a different linear combination, instead of the two equations in (\ref{IMC}), one can use

\begin{eqnarray}
[ \cR ]_{\pm} &=& 0  \nonumber\\
\left[  \Phi  \right]_{\pm} & = & 0 \label{AMC}
\end{eqnarray}

It is easy to write these variables in terms of the Mukhanov variable $v$, for which there is a simple analytical solution \cite{Contaldi:2003zv} in the two phases of interest:

\begin{eqnarray*}
\cR & = & \frac{v}{z} \\
\Delta \Phi  & =&  \frac{1}{2} \frac{{\phi'_{0}}^2}{\cH} \left( \frac{v}{z} \right)' \\
z & = & \frac{a \phi'_0}{\cH}
\end{eqnarray*}
Indeed for

\begin{eqnarray*}
&\mbox{For}& \hspace{3ex} \eta < 0 \hspace{5ex} v_k(\eta) = \sqrt{\frac{\pi}{8 H}} \,\,\sqrt{1 + 2 H \eta} \,\,H_0^{(2)}\left( k \eta + \frac{k}{2 H}\right)  \\ \\
&\mbox{For}& \hspace{3ex} \eta > 0 \hspace{5ex} v_k(\eta) = C e^{-i ( k \eta - k/H)} \left( 1 - \frac{i}{k \eta- k/H}\right) + D e^{i ( k \eta - k/H) }\left( 1 + \frac{i}{k \eta- k/H}\right)
\end{eqnarray*}
There are only two arbitrary constants $C$ and $D$, that will be fixed by the matching conditions (\ref{AMC}). They turn out to be:

\begin{eqnarray}
C &=& \frac{e^{- i k / H}}{\sqrt{96 H /\pi}} \left[ \sqrt{\epsilon} \,\, H_0^{(2)} \left( \frac{k}{2 H} \right) - \frac{3}{\sqrt{\epsilon}}  \left( \frac{H}{k} + i \right) H_1^{(2)} \left( \frac{k}{2 H} \right) \right] \nonumber \\  \nonumber\\
D &=& \frac{e^{ i k / H}}{\sqrt{96 H /\pi}} \left[ \sqrt{\epsilon} \,\, H_0^{(2)} \left( \frac{k}{2 H} \right) - \frac{3}{\sqrt{\epsilon}}  \left( \frac{H}{k} - i \right) H_1^{(2)} \left( \frac{k}{2 H} \right) \right]  \label{CD}
\end{eqnarray}
where $\epsilon= \frac{1}{2} \frac{{\phi'_0}^2}{H^2}$. The power spectrum for modes outside the horizon is:

$$ \cP_{\cR}(k)= \frac{H^2}{{\dot{\phi_0}}^2} \frac{H^2}{k^2} |C-D|^2 $$ It is easy to see that in the limit $k/H \gg 1$, that is for modes that were inside the horizon when inflation started:

$$ |C-D|^2 \sim \frac{3}{ 2 \epsilon} \frac{1}{k} \cos^2\left(\frac{k}{H} \right) $$ which gives

$$\cP_{\cR}(k)= \frac{3}{4 \, \epsilon^2 } \frac{H^2}{k^3} \cos^2\left(\frac{k}{H} \right) $$
This completely agrees with the result obtained using the Schr\"{o}dinger approximation (\ref{fr}).

\section{Numerical Results}\label{numericalresults}
Although the assumptions made in the above analysis are self-consistent and appear robust, it seems prudent to confirm the results with a numerical study. We do this by studying the production of perturbations during inflation with a continuous potential close to a step function:
\begin{equation}
V=\frac{\Lambda^4}{2}(1+\tanh\left(f\phi\right))\left(1-\alpha\phi\right)\, .
\end{equation}
This potential is identical to the one studied analytically above, in the limit $f\rightarrow\infty$. Our goal in studying this potential is not to completely analyse a particular model, but rather to ensure that the assumptions in the previous analysis are reasonable, and that the transition between sudden and adiabatic regimes has a measurable effect.

\begin{figure}
\centering
\includegraphics[width=\textwidth]{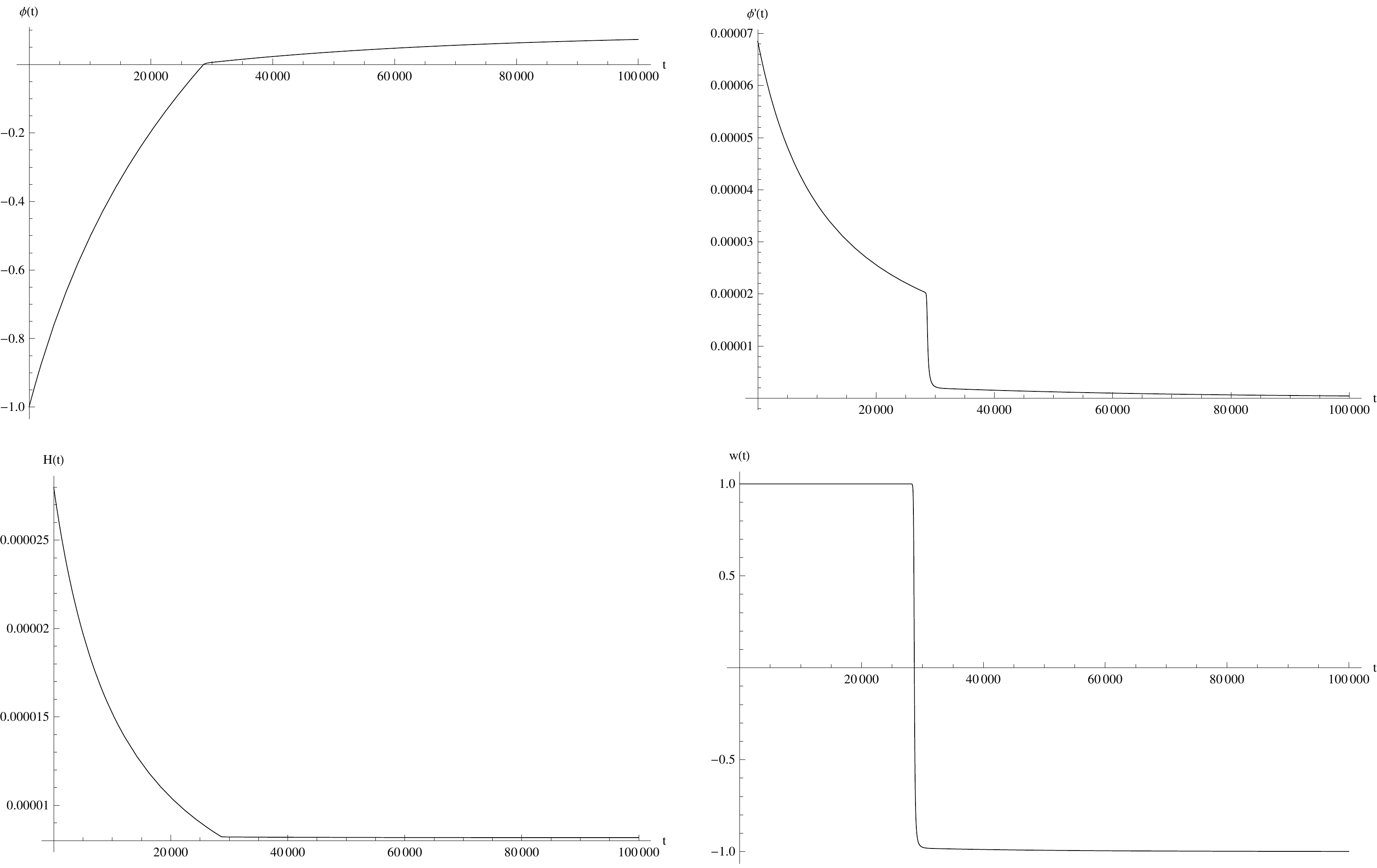}
\caption{Plots showing the evolution of background field quantities as the system transitions int inflation. The top row has $\phi$ and $\dot{\phi}$, the bottom $H$ and $w$.}
\label{tanhbackground1}
\end{figure}

It is straightforward\footnote{All numerical differential equation solving has been done with Mathematica NDSolve and all interpolation with Mathematica Interpolation.} to solve the background equations of motion in comoving time\footnote{In order to achieve a numerically stable solution it is necessary to use different time coordinates during the early and late stages of the field evolution. Once deep in the inflationary regime, the number of e-foldings is the time coordinate of choice.} during the onset of inflation. In figure \ref{tanhbackground1} the evolution of $\phi$, $\dot{\phi}$, $H$ and $w$ are shown as the field evolves past the step and enters the inflationary regime. Note the sharp drop in the field velocity that suggests a slow roll phase has begun. The initial conditions have been tuned in order to ensure the transition from kinetic to potential domination is suitably rapid. This can be seen via evolution of $w$, which appears to indicate a sharp change in the equation of state, thus putting us in a regime where the sudden approximation used in the previous section should be appropriate for some set of modes.

\begin{figure}
\centering
\includegraphics[width=\textwidth]{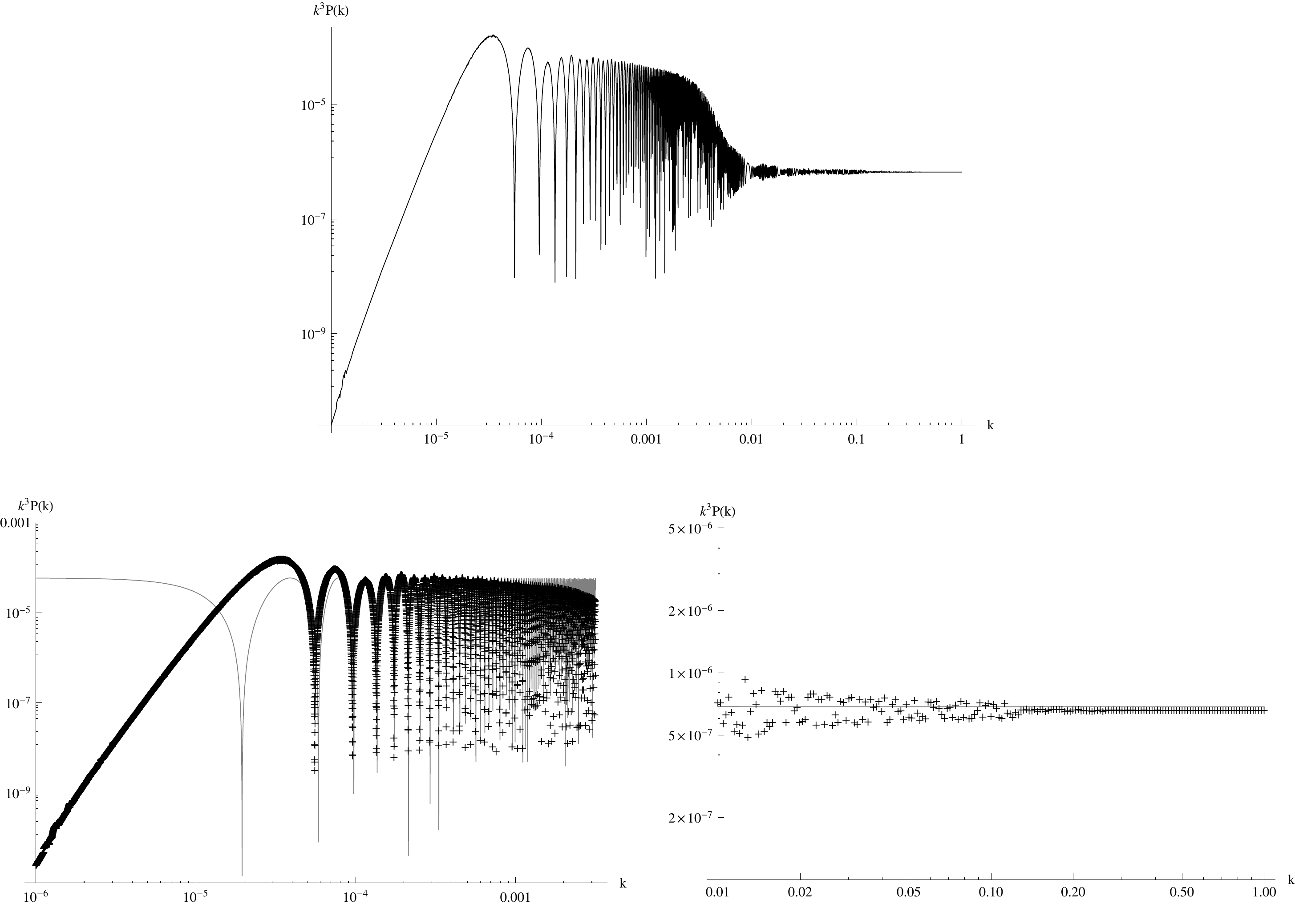}
\caption{These plots show the power spectrum generated from solving the Schr\"{o}dinger equation as described in the text. At the top s the full spectrum from $k=10^{-6}$ to $k=1$, with the spectrum reconstructed using Mathematica's Interpolation routine. At the bottom left we show points from power spectrum in the range $k=10^{-6}$ to $k=10^{-2.5}$ along with a fit to $c(k)\left(\frac{H}{\dot{phi}}\right)^{\beta(k)}\frac{H^2}{k^3}\cos^2\left(k/a_iH_i\right)$ ($c(k)=2.01$, $\beta(k)=2.83$). And at the bottom right the spectrum is shown in the range $k=10^{-2}$ to $k=1$, with a fit to the standard, scale invariant result.}
\label{numericalpower}
\end{figure}

Armed with the numerical background solution, our next step is to solve the Schr\"{o}edinger equation for each fourier modes as an initial value problem, using the initial conditions discussed in section \ref{initialcond}. The Schr\"{o}edinger equation in momentum space is:
\begin{equation}
i\partial_t\psi_k=\left(-\frac{2}{a z^2}\frac{\partial^2}{\partial R_k\partial R_{-k}}+\frac{k^2z^2}{2a}R_kR_{-k}\right)\psi_k\, .
\end{equation}
As previously we assume a Gaussian wavefunction so that $\psi_k$ has the form:
\begin{equation}
\psi_k=\exp\left[-\frac{f_k(t)}{2}R_kR_{-k}\right]\, .
\end{equation}
This leads to the following (Riccati) equation for $f_k$:
\begin{equation}
\partial_tf_k=ik^2\frac{z^2}{a}-i\frac{f_k^2}{az^2}\, .
\end{equation}
We solve this equations using the initial conditions for a kinetic dominated background given in equation \ref{abdv}:
\begin{equation}
\Re F (t) + i \, \Im F(t) = i \,  m(t) \omega(t) \, \frac{ H_1^{(1)}( w_0/2) }{H_0^{(1)} (w_0/2)}\, .
\end{equation}
$w_0 = k/a_0H_0$ is the product of the co-moving wavenumber and horizon at some initial time before inflation starts. The resulting power spectrum (for $\Lambda^4=10^{-10}m_p^4$, $\alpha=10^{-2}$, $f=500m_p^{-1}$, $\phi(0)=-1$ and $\dot{\phi}(0)=6.8505\times10^{-9}m_p^2$) is given in figure \ref{numericalpower}. For sufficiently small $k$, we see that the power spectrum is well approximated by a function of the form following from equation \ref{k3d}:
\begin{equation}
P(k)=c(k)\left(\frac{H}{\dot{\phi}}\right)^{\beta(k)}\frac{H^2}{k^3}\cos^2(w_i)\, .
\end{equation}
For $k$ in the range $10^{-4}$ to $10^{-3}$, we have a good fit for $\beta(k)$ close to 3 and $c(k)$ close to 2. For larger values of $k$ the power spectrum is closer to the result one would expect for an adiabatic transition to the inflationary regime, with the conventional result:
\begin{equation}
P(k)=\left(\frac{H}{\dot{\phi}}\right)^2\frac{H^2}{k^3}\, .
\end{equation}

While the above appears to support the analytical results found above, there are several observations that should be made. Firstly, the effect seen is the change in amplitude is somewhat less dramatic than the purely sudden transition. This appears to be a result of the finite transition time from kinetic to potential domination; as one adjusts parameters, one finds that the slower the transition, the less pronounced the change in amplitude. However, even though a realistic model may not have the $\mathcal{O}(1/\epsilon)$ change in amplitude between the sudden and adiabatic regimes, we still see a change of order $1/\sqrt{\epsilon}$ with minimal fine-tuning. The critical value of $k$ demarcating the switch from sudden to adiabatic modes is related to the second derivative of $z$. In conformal time, the equation motion of the Mukhanov variable is:
\begin{equation}
\frac{v''}{v} + ( k^2-\frac{z''}{z} ) =0
\end{equation}
As long as $\frac{z''}{z}$ is finite, the solution for $k^2 \gg \frac{z''}{z}$ won't have information about the transition.

We also note that for scales larger than the horizon ($k/aH\ll1$, which corresponds to $k\ll5\times10^{-5}$) there is an unsurprising drop off in power. This is due to the fact the gaussian spread for these modes ($f_k(t)$) will not evolve during inflation, and so remains at the flat value set by the initial conditions giving a $1/k$ power spectrum. Of course, as discussed in the body of the text, these long wavelength modes are precisely those whose initial conditions are indeterminate -- there is no requirement from renormalization of the energy momentum tensor that they have the form that we must impose on shorter wavelength modes.

Finally, the numerical model illustrated here does have some fine-tuning, roughly of order $10^{-5}$ in the initial field velocity. While this is not insubstantial, neither is it particularly egregious by the standards of inflationary model building. As we point out in the conclusions, this is consistent with our view that the effect discussed in this paper is not impossible to see given appropriate criteria are met.

\bibliographystyle{JHEP}
\bibliography{initialstate}

\end{document}